\documentclass[journal=jctc,manuscript=article]{achemso}

\usepackage{graphicx}% Include figure files
\usepackage{subcaption}
\usepackage{dcolumn}% Align table columns on decimal point
\usepackage{bm}% bold math
\usepackage{multirow}% Added by H. Shin to plot multi-row table
\usepackage{gensymb}% Added by Tomas. For the Celsius symbol
\usepackage{amsmath}
\usepackage{placeins} % FloatBarrier

\usepackage{xcolor}

\title{Systematic Improvement of Quantum Monte Carlo Calculations in Transition Metal Oxides: sCI-Driven Wavefunction Optimization for Reliable Band Gap prediction}

\author{Hyeondeok Shin}
 \email{hshin@anl.gov}
\affiliation{Computational Science Division, Argonne National Laboratory, Argonne, Illinois 60439, USA}
\author{Kevin Gasperich}
 \email{kgasperich@anl.gov}
\affiliation{Computational Science Division, Argonne National Laboratory, Argonne, Illinois 60439, USA}
\author{Tomas Rojas}
\affiliation{Department of Chemical  Engineering,  University of Illinois at Chicago, Chicago, Illinois 60608,  United States}
\author{Anh T. Ngo}
\affiliation{Department of Chemical  Engineering,  University of Illinois at Chicago, Chicago, Illinois 60608,  United States}
\affiliation{Material Science Division, Argonne National Laboratory, Argonne, Illinois 60439, USA}
\author{Jaron T. Krogel}
\email{krogeljt@ornl.gov}
\affiliation{Materials Science and Technology Division, Oak Ridge National Laboratory, Oak Ridge, TN 37831, United States}

\author{Anouar Benali}
\email{benali@anl.gov}
\affiliation{Computational Science Division, Argonne National Laboratory, Argonne, IL 60439, United States}

\date{\today}% It is always \today, today,
\begin{document}

\begin{abstract}

Accurate determination of electronic properties of correlated oxides remains a significant challenge for computational theory. Traditional Hubbard-corrected density functional theory (DFT+U) frequently encounters limitations in precisely capturing electron correlation, particularly when predicting band gaps. We introduce a systematic methodology to enhance the accuracy of diffusion Monte Carlo (DMC) simulations for both ground and excited states, focusing on LiCoO$_2$ as a case study. By employing a selected CI (sCI) approach, we demonstrate the capability to optimize wavefunctions beyond the constraints of single-reference DFT+U trial wavefunctions.
We show that the sCI framework enables accurate prediction of band gaps in LiCoO$_2$, closely aligning with experimental values and substantially improving upon traditional computational methods. The study uncovers a nuanced mixed state of $t_{2g}$ a $e_g$ orbitals at the band edges that is not captured by conventional single-reference methods, further elucidating the limitations of PBE+U in describing $d$-$d$ excitations.
Our findings advocate for the adoption of beyond-DFT methodologies, such as sCI, to capture the essential physics of excited state wavefunctions in strongly correlated materials. The improved accuracy in band gap predictions and the ability to generate more reliable trial wavefunctions for DMC calculations underscore the potential of this approach for broader applications in the study of correlated oxides. This work not only provides a pathway for more accurate simulations of electronic structures in complex materials but also suggests a framework for future investigations into the excited states of other challenging systems.

\end{abstract}

%\keywords{quantum Monte Carlo method, density functional theory}
%\maketitle

%\tableofcontents

\section{\label{sec:level1}Introduction}
Quantum Monte Carlo (QMC)\cite{Reynolds_1982} is a class of computational methods solving the many-body Schr\"odinger equation stochastically. By explicitly taking into account electron-electron interactions, QMC can accurately describe
weakly- to strongly-correlated materials, with a limited and controlled number of approximations. Recent years have seen a significant increase in development and application of the fixed-node diffusion Monte Carlo (DMC)\cite{Reynolds1990,foulkes01} method for its capacity at recovering, in a systematic and improvable manner, properties of ground and excited states of a wide variety of materials and molecules, while providing, not without challenges, ground-breaking results for transition metal (TM) oxides and layered materials.~\cite{foyevtsova14,shin17,shin2017,kylanpaa17,shin18,Morales2012,Benali_2014,benali16,Kolorenvc_2011,Dubecky_2016}

Applications of DMC to fermionic systems imposes an antisymmetric trial wavefunction implicitly defining the nodal surface constraining the projection and sampling.
If the nodes of the trial wavefunction are exact, then the DMC energy will converge to the exact ground state energy of the system.  Any deviation from the exact nodal surface will cause an increase in the calculated DMC energy, referred to as the \textit{fixed-node} (FN) error \cite{Anderson1980}. The DMC energy thus provides a variational upper bound to the exact ground state energy.

This variational property of DMC makes it possible to improve the nodal surface by simply evaluating the DMC energy using several different nodal surfaces and choosing the lowest one.
Mean-field methods such as Hartree-Fock (HF)\cite{Hartree1928,Fock1930,Slater1930,Hartree1935} and density functional theory (DFT)\cite{Hohenberg_1964,Kohn_1965,ParrBook} provide a relatively inexpensive way to generate trial wavefunctions, but these methods don't directly optimize the wavefunction nodes, and the nodal surface quality from DFT depends strongly on the choice of exchange-correlation (XC) functional used.~\cite{shin17,Shin2021}

In the case of strongly-correlated materials such as TM oxides and complexes, it is common to use a single determinant of DFT orbitals to fix the position of the nodes. The use of DFT+U orbitals with a Hubbard U correction\cite{anisimov91} to take into account on-site Coulomb interaction for localized $d$ orbitals on the TM atoms\cite{dudarev98} has been applied to various TM oxides, and the resulting DMC energies have been in excellent agreement with experimental values.~\cite{foyevtsova14,shin17,shin2017,kylanpaa17,shin18,Zheng2015,Busemeyer2019,Yu2017,Benali2016} 

Optimizing the nodal surface for excited states presents more challenges than for the ground state.
While DFT or DFT+U orbitals can provide a suitable nodal surface for the ground state, optimizing the nodal surface for the excited state in the same way as the ground state is controversial.~\cite{shin17} 
Many DMC results using DFT orbitals have claimed that DMC tends to overestimate band gaps compared to the experimental results for various correlated materials.~\cite{shin17,hunt18,hunt20,nikaido21}
Accurate prediction of band gaps is fundamental and plays a pivotal role in a myriad of scientific and technological applications, ranging from semi-conductors to energy storage systems, making the task of identifying a set of independent orbitals that accurately represent both the ground and excited states  particulary valuable.
The goal of this work is to introduce an \textit{ab initio} method that accurately describes the orbitals, and consequently the properties, of both the ground and excited states of complex materials. This is achieved by going beyond traditional DFT or DFT+U orbitals, employing multi-Slater determinants trial wavefunctions that more effectively capture the complex electron correlation often present in the $d$-manifold of transition metal oxides.

Beyond the use of single Slater determinant trial wavefunctions, the use of multi-Slater determinants has proven to systematically reduce the FN error,\cite{Morales2012,Caffarel_2016a,Giner_2013,Scemama2016,Caffarel_2016b,Townsend2020,Deible2015,Filippi2016,Harkless2006,Sun1992,Brown2007,Filippi2022} but this approach has been mainly restricted to small molecules due to a significant increase in the computational cost of the DMC calculations.~\cite{Clark2011,Giner_2013}
Recent efficient implementations of selected configuration interaction (sCI) methods\cite{Huron_1973,Dash_2018,Dash_2019,Garniron_2019b,Holmes_2016,Sharma_2017,Holmes_2017,Chien_2018,Cimiraglia_1987,Evangelisti_1983,Evangelista_2014} have made it possible to generate multideterminant wavefunctions without restricting the excitations to only include chemically relevant orbitals, so no prior knowledge of the system is required. sCI does not remove the exponential scaling of the size of the full configurational space, but it increases the size of system that can be studied before the cost becomes prohibitive. Moreover, sCI has also been shown to be effective at generating multideterminant trial functions for excited states, and it is possible to perform sCI in such a way as to balance the errors of the ground and excited states to obtain accurate energy differences.~\cite{Scemama_2018b,Loos_2018b,Loos_2020b,Loos_2020a,Scemama_2019,Loos_2019,Neuscamman2020,neuscamman2019,Garner2020,Filippi2022}
More recently, the use of multideterminants has been extended to periodic systems,\cite{Benali2020,Malone2020} and in the case of GeSe\cite{Shin2021}, multideterminant wavefunctions were used on a small cell to assess the accuracy of the nodal surface from a single determinant trial wavefunction.
Building on the advancements in sCI methods for solids, this work further explores their potential in accurately modeling complex materials. While sCI significantly enhances our ability to generate wavefunctions that capture both ground and excited states without prior system knowledge, the intrinsic complexity and size of the systems of interest limit full convergence within the sCI determinant space. Recognizing this, our approach leverages the generation of natural orbitals (NOs) from the most comprehensive determinant expansion achievable independently for both ground and excited states. This iterative process, aimed at achieving energy convergence, effectively mitigates the computational challenges associated with approximating the Full Configuration Interaction (FCI) limit. Consequently, it enables a balanced description of electronic states, overcoming the limitations of DFT in characterizing excited states as accurately as ground states. This approach not only refines the trial wavefunctions for DMC calculations but also ensures a more accurate representation for both ground and excited states.\\

\begin{figure}[t]
 \includegraphics[width=6 in]{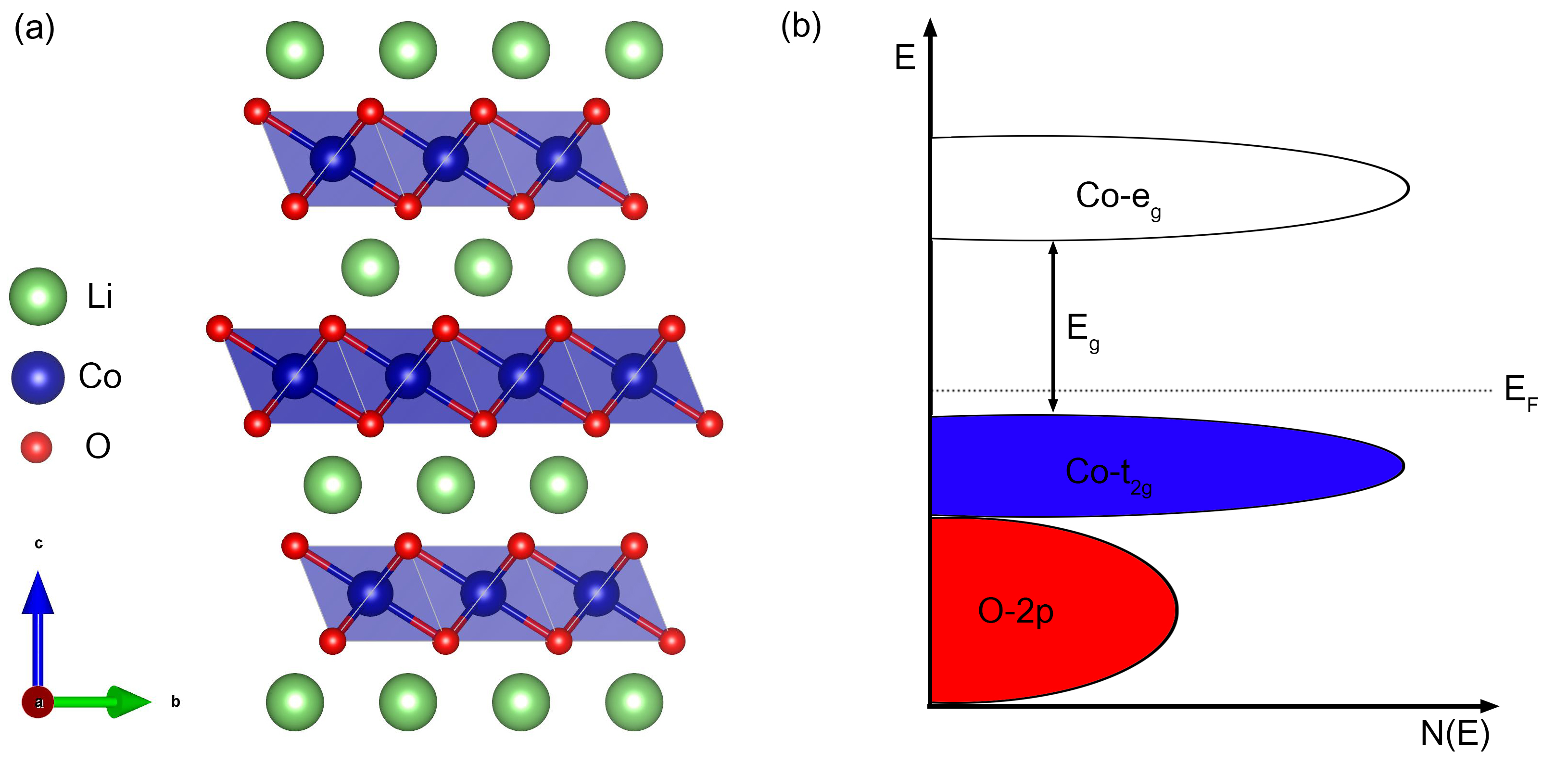}
 \caption{(a) R-3m rhombohedral structure of bulk LiCoO$_2$. (b) Cartoon of idealized electron density of states for LiCoO$_2$ based on a fully ionic picture.}
 \label{fig:LiCoO2}
\end{figure}

In this study, we investigate in detail the properties of trial wavefunctions used for ground and excited states of a strongly-correlated transition metal oxide for which the DMC band-gaps with a single Slater determinant were found to overestimate the experimental value.~\cite{schiller15,shin17}
As a symptomatic material, we choose to apply our investigative tools to analyze the energy band-gap of lithium cobalt dioxide (LiCoO$_2$). LiCoO$_2$ was the cathode material used in the first commercialized lithium-ion battery, and since then it has been used extensively as the active material in the positive electrode.~\cite{Blomgren2017} The crystal structure of LiCoO$_2$ forms an ordered rocksalt phase, a layered phase with R-3m symmetry (see Fig.~\ref{fig:LiCoO2}(a)).~\cite{Mizushima1980,Park2010}

Due to high interest in LiCoO$_2$ as a cathode material for rechargeable batteries, it has been reliably characterized by DFT in many ways, although there are crucial deficiencies in its description. While DFT+U is able to recover some properties, it presents several problems that introduce limitations within the simulations, among them, underestimating the voltage by approximately 0.3 V\cite{Chevrier2010}, as well as overestimating Li order/disorder transition temperatures by up to one hundred degrees.~\cite{VanderVen1998,Isaacs2017a} Additionally, DFT+U seems to show a transition to a high-spin state\cite{elp91} not observed in experiments of LiCoO$_2$,\cite{Menetrier1999} which is in stark contradiction with measurements.~\cite{Pourovskii2007} Recent work combining DFT with dynamic mean-field theory (DMFT)\cite{Kotliar06} claimed that occupancies of both $t_{2g}$ and $e_{g}$ states in LiCoO$_2$ are dependent on the value of U,\cite{issacs2020} leading the authors to conclude that finding the optimal value of U for LiCoO$_2$ is important in determining the exact distribution of $t_{2g}$ and $e_{g}$ populations at the valence band maximum (VBM) and conduction band minimum (CBM) that are nominally expected to be separated as a low spin state ($t_{2g}^{6}e_{g}^{0}$ in a fully ionic picture) as seen in Fig.~\ref{fig:LiCoO2}(b).~\cite{laubach09}
Using single determinant DMC (SD-DMC) with a DFT+U trial wavefunction and optimizing the value of U by leveraging the variational principle inherent in DMC revealed a distinctive outcome. Rather than converging towards a U value akin to those typically used in DFT calculations,  the ground state of LiCoO$_2$ necessitated an exceptionally high value of U, exceeding 10 eV. Additionally, the DMC-calculated optical gaps were found to be significant overestimates. These results highlight the limitations of DFT and DFT+U orbitals in accurately reflecting the characteristics of both the ground and excited states. They also point to a markedly pronounced fixed-node (FN) error in either state when utilizing trial wavefunctions based on DFT, indicating that conventional methods may fall short in accurately simulating the complex electronic structure of materials such as LiCoO$_2$.\\

We have implemented an alternative scheme for obtaining trial wavefunctions for LiCoO$_2$ which is based on sCI.  With a large enough space of determinants, we can capture much of the correlation energy in both the ground and excited states, allowing us to interrogate both the essential nature of the low energy states in LiCoO$_2$ and the quality of the DFT+U approximation in describing them.
Our multi-reference calculations show that sCI natural orbitals offer improved nodal surfaces for LiCoO$_2$'s excited state compared to DFT-based orbitals. Furthermore, our comprehensive excitation analysis reveals that the majority of $\Gamma$-point excitation is particle-hole type, while a non-negligible portion displays mixed occupation between $t_{2g}$ and $e_{g}$ states near the Fermi level, challenging mean-field theories in representing its spectral properties.

\section{\label{sec:level2}Computational details}
All QMC calculations were carried out using the FN-DMC method as implemented in the \textsc{QMCPACK} code.~\cite{QMCPACK} Single- and multi-determinant calculations used Slater-Jastrow trial wavefunctions where dynamic correlation were described by one- two- and three-body Jastrow functions. The parameters of these Jastrow factors were optimized independently for each trial wavefunction via the linear method\cite{10.1103/PhysRevLett.98.110201} within variational Monte Carlo (VMC). To remove the effect of high-energy core electrons and simplify valence wavefunctions, all calculations used norm-conserving correlation consistent effective core potentials (ccECPs)\cite{bennett17,wang19} (plane-wave basis sets) and energy-consistent small-core potentials (Gaussian basis sets) developed by Burkatzki, Filippi, and Dolg (BFD).~\cite{burkatzki07,burkatzki08} The single-particle orbitals in the Slater determinants were obtained within the Kohn-Sham DFT+U scheme based using both plane-wave and Gaussian basis sets.

\subsection{\label{subsec:comp-details-singleref}Single-reference DFT-based trial wavefunction calculations}
Single Slater determinants using plane-wave basis sets were generated with the \textsc{quantum espresso} package\cite{giannozzi09} for bulk LiCoO$_2$ within a 12 atom unit cell.  Calculations were performed with fully converged Monkhorst-Pack k-point grids of size ${6\times6\times6}$.~\cite{Monkhorst_1976}
We chose lattice parameters for the R-3m rhombohedral structure of bulk LiCoO$_2$ (see Fig.~\ref{fig:LiCoO2}(a)) as ${a = 2.83 \text{ \AA}}$ and ${c = 11.41 \text{ \AA}}$, which were experimentally determined using X-ray diffraction.~\cite{reimers92,laubach09}
In this study, we used norm-conserving ccECPs with 700 Ry kinetic energy cut-off for the Co atom.~\cite{bennett17,wang19}
We considered PBE and local density approximation (LDA) XC functionals, and a Hubbard U correction\cite{anisimov91} was employed to take on-site Coulomb interaction into account for localized $d$ orbitals on the Co atoms.~\cite{dudarev98}

DMC calculations for the single-reference trial functions in this study were performed with a  0.005 Ha$^{-1}$ time step and size-consistent $T$-moves.~\cite{casula10}
In order to minimize the finite-size effects in the solid, DMC total energies were averaged over the maximum 125 twists (twist-averaged boundary conditions)\cite{lin01} with addition of the model periodic Coulomb (MPC) interaction\cite{drummond08} and Chiesa-Ceperley-Martin-Holzmann kinetic energy correction.~\cite{chiesa06}
In addition, we computed the corrected DMC energy at various supercell sizes and extrapolated these to the bulk limit in order to estimate the DMC energy at the thermodynamic limit. 
DMC calculations were performed for single-determinant DFT wavefunctions obtained with a range of U values. These were completed with supercells consisting of 48 atoms (360 electrons) in a 2x2x1 tiling of the original 12-atom cell, with a total of 64 supercell twists.

\subsection{Multi-reference sCI wavefunction calculations}
For the multideterminant trial wavefunctions, electronic integrals and Gaussian orbitals were obtained from \textsc{pyscf}\cite{PySCF} using BFD pseudopotentials\cite{burkatzki07,burkatzki08} and corresponding triple-zeta Gaussian basis sets. While we used the same simulation cells as in the plane-wave calculations, consisting of 12 atoms per cell, due to the cost of the simulations, only the $\Gamma$ point was considered with Gaussian orbitals. Due to the large number of orbitals in the system (405), similar to our previous work on multideterminant bulk diamond\cite{Benali2020}, we retained the 200 lowest-energy orbitals and performed subsequent sCI calculations in the space of these orbitals. Compared to the plane-wave basis set calculations, the use of different pseudopotentials and a truncated Gaussian basis set lead to the same single determinant DMC energy (agreeing to $3\pm5$ mHa), justifying these approximations. 

As an sCI method, we use the configuration interaction using a perturbative selection made iteratively (CIPSI) algorithm, \cite{Huron_1973} as implemented in the \textsc{quantum package}  code \cite{Garniron_2019b}.
CIPSI produces an accurate wavefunction expanded in a basis of Slater determinants.  As the convergence rate toward the FCI limit can be improved with better orbital sets, we considered Slater determinant spaces constructed from the KS orbitals produced by PBE+U calculations with several values of U (U=0, 3, 5, 7, and 11 eV).
After performing two independent CIPSI calculations for the ground state and the first excited state in these PBE+U orbitals, we used the density matrices from the  multideterminant expansions to generate a set of independent natural orbitals (NOs) for the ground and excited state. We performed a subsequent CIPSI calculation in this first set of NOs and repeated this until reaching convergence of the variational energy at a fixed number of determinants.  This was achieved after a total of 5 repetitions to obtain the final set of NOs (see section~\ref{sec:CIPSI_conv} for more details).

\section{\label{sec:results}Results}
We obtained an optimal value of U for LiCoO$_2$ by minimizing the DMC energy with respect to U.
We then analyze the results of DFT/DFT+U and DMC calculations with our optimized U values and with typical U values that have been used in previous studies.
We compare these to CIPSI results using a several single- and multi-determinant wavefunctions obtained from different sets of starting orbitals.

\subsection{\label{subsec:results-singleref}Single-reference DMC results}
\begin{figure}[t]
 \includegraphics[width=6 in]{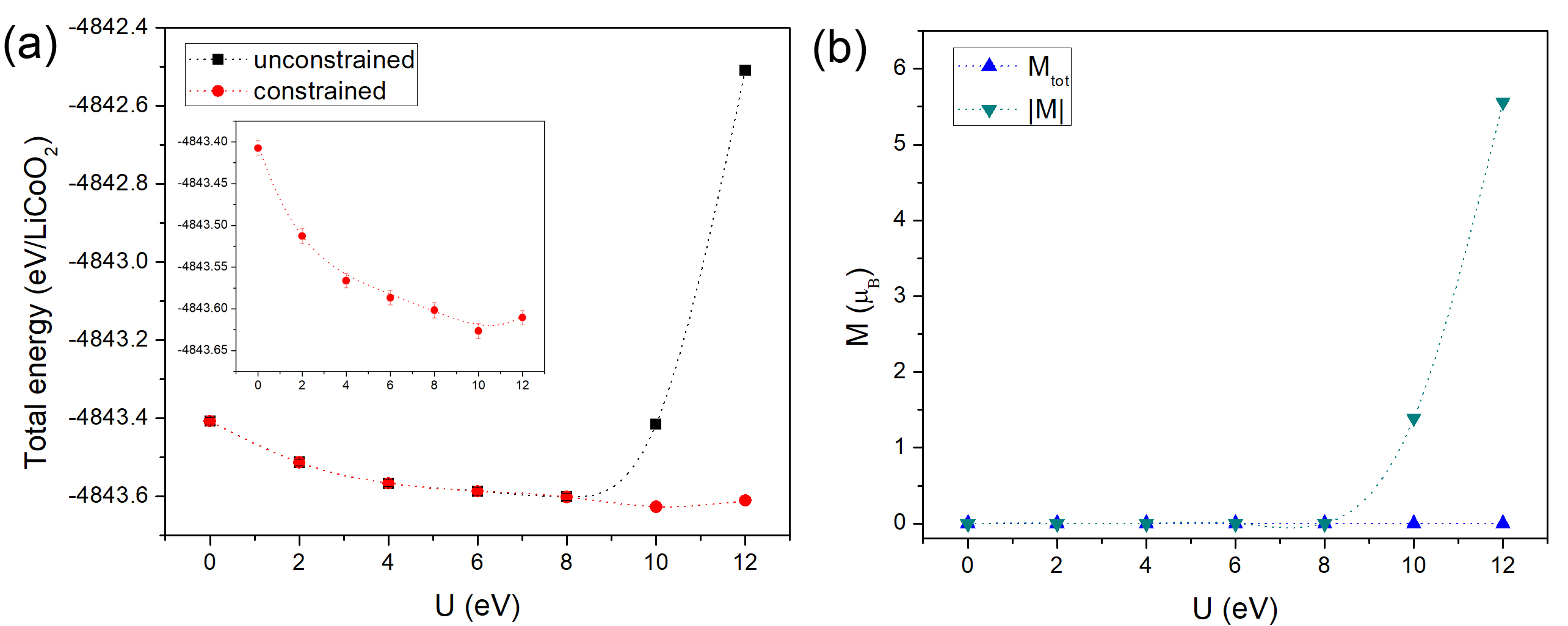}
 \caption{(a) DMC total energy as function of U in the PBE+U wavefunction with and without the constraint of zero magnetization of the Co atoms.  Inset figure represents smaller y-axis scale of DMC total energy with the constraint.
 (b) PBE+U total and absolute magnetization of LiCoO$_2$ without the constraints as function U. Dotted lines are a guide for eyes.
 }
 \label{fig:DMC_U}
\end{figure}

An appropriate value of U for DFT+U is often chosen by the linear-response approach, which is an effective self-consistent method for finding the optimal value of U within the Kohn-Sham scheme.~\cite{cococcioni05}      
Previous DFT+U studies for LiCoO$_2$ used values of U in the range of 3.0 - 3.3 eV,\cite{dixit16,Chakraborty2018} obtained empirically via fitting the DFT+U oxidation energy of CoO to experimental data.~\cite{Wang2006} 
As mentioned above, DMC is a variational method whose accuracy depends solely on the quality of the nodal surface, which is fixed by the Slater determinant in the trial wavefunction. In the case of a DFT+U Slater determinant, the value of the U parameter can be used as variational \textit{knob} to optimize the nodal surface in a simple 1-D scan minimizing the DMC total energy. This selection of the value of the U parameter through DMC minimization has been very successful at describing ground state properties of many transition metal oxides.~\cite{shin17,shin18}  
However, applying this method to LiCoO$_2$ yields a different outcome; figure~\ref{fig:DMC_U}(a) shows DMC total energy as function of Hubbard U for PBE+U trial wavefunction. 

We here consider two separate sets of PBE+U orbitals, one with unconstrained magnetization and one constrained to have zero magnetization.
This constraint was imposed because the unconstrained PBE+U calculation gives positive absolute magnetization with the opposite sign of a magnetic moment between Co and O atoms beyond U = 10 eV (see Fig.~\ref{fig:DMC_U}(b)).
Although the non-magnetic system has a higher total energy than the anti-ferromagnetism in PBE+U with large U (see Supplemental Information.),\cite{SI} it is worthwhile to compare DMC total energies using both non-magnetic and anti-ferromagnetic trial wavefunctions in order to determine which magnetic ordering will provide a better nodal surface (i.e. a lower variational DMC energy).
In Figure~\ref{fig:DMC_U}(a), both DMC total energies are almost identical below U = 10 eV, where the unconstrained PBE+U had converged with zero magnetization ($M=0$). 
At U = 10 eV, where the unconstrained PBE+U no longer yields a non-magnetic state ($M = 1.3\,\mu_{B}$), the constrained PBE+U orbitals ($M = 0\,\mu_{B}$) yield a lower DMC energy---and thus have a better nodal surface---than the unconstrained ones.
Bulk LiCoO$_2$ has been known to possess non-magnetic properties with the low-spin state where all valence electrons of the Co$^{+3}$ ion reside in a fully occupied $t_{2g}$ shell,\cite{laubach09} so the observed positive absolute magnetization at large values of U represents a departure from the low-spin state.

The optimal value of U within a DMC framework can be evaluated by fitting a quartic function of U to the DMC total energies obtained using several DFT+U trial wavefunctions (using the same XC functional with varying value of U).~\cite{shin17,shin18} 
The inset in Figure ~\ref{fig:DMC_U}(a) omits the energies obtained from the unconstrained procedure to more clearly show the effect on the DMC energy of varying U.
As U increases from 0, the DMC total energy rapidly decreases due to the improvement in the nodal surface of the PBE+U wavefunction caused by the Hubbard U correction.
Using a quartic fit to DMC total energy as function of U, the optimal value of U (U$_{opt}$) for PBE+U wave function is estimated to be 11.0(4) eV.
Note that LDA+U and PBE+U wavefunctions provide almost identical DMC total energies at the PBE U$_{opt}$, but the LDA+U optimal U value is 13.8(2) eV (see Supplemental Information).~\cite{SI}
We estimate the cohesive energy of LiCoO$_2$ with full consideration of the two-body finite-size effect by extrapolating the DMC total energy to the bulk limit using various sizes of supercell, consisting of 9, 12 and 24 formula units of LiCoO$_2$ (details of the finite-size analysis can be found in the Supplemental Information).~\cite{SI}
In the bulk limit, the DMC cohesive energy for bulk LiCoO$_2$ is estimated to be 18.03(3) eV/f.u. while PBE and LDA cohesive energies are 17.73 and 22.43 eV/f.u., respectively.
The experimental value of the cohesive energy of bulk LiCo$_2$ is estimated to be 18.22 eV by using the cohesive energies of components in the equation for the reaction, 0.5Li$_2$O + CoO + 0.25O$_2$ = LiCoO$_2$.~\cite{Jog1985,lide95,Glasser2016}
Details of the calculation of the cohesive energy can be found in the Supplemental information.
Despite the unusually large value of U$_{opt}$, the near agreement of the DMC cohesive energy (18.03(3) eV/f.u.) with the experimental value (18.22 eV/f.u.) suggests that the PBE+U wavefunction provides a good nodal surface for the ground state of LiCoO$_2$.
A low-spin state for LiCoO$_2$ is expected to show fully-occupied 3$d$-t$_{2g}$ and empty 3$d$-e$_g$ states separated into the valence and conduction bands.
In order to demonstrate the actual effect of the Hubbard repulsion U on splitting the 3$d$ orbitals into upper and lower Hubbard bands in DFT, we compute a projected density of states (PDOS) of LiCoO$_2$ using PBE and PBE+U.
In order to obtain the populations of $t_{2g}$ and $e_{g}$ states on Co$^{+3}$ in LiCoO$_2$, the PDOS is calculated in the local coordinate frame where the atomic occupation matrix is diagonal.~\cite{Mahajan2021}
\begin{figure}[t]
\includegraphics[width=6 in]{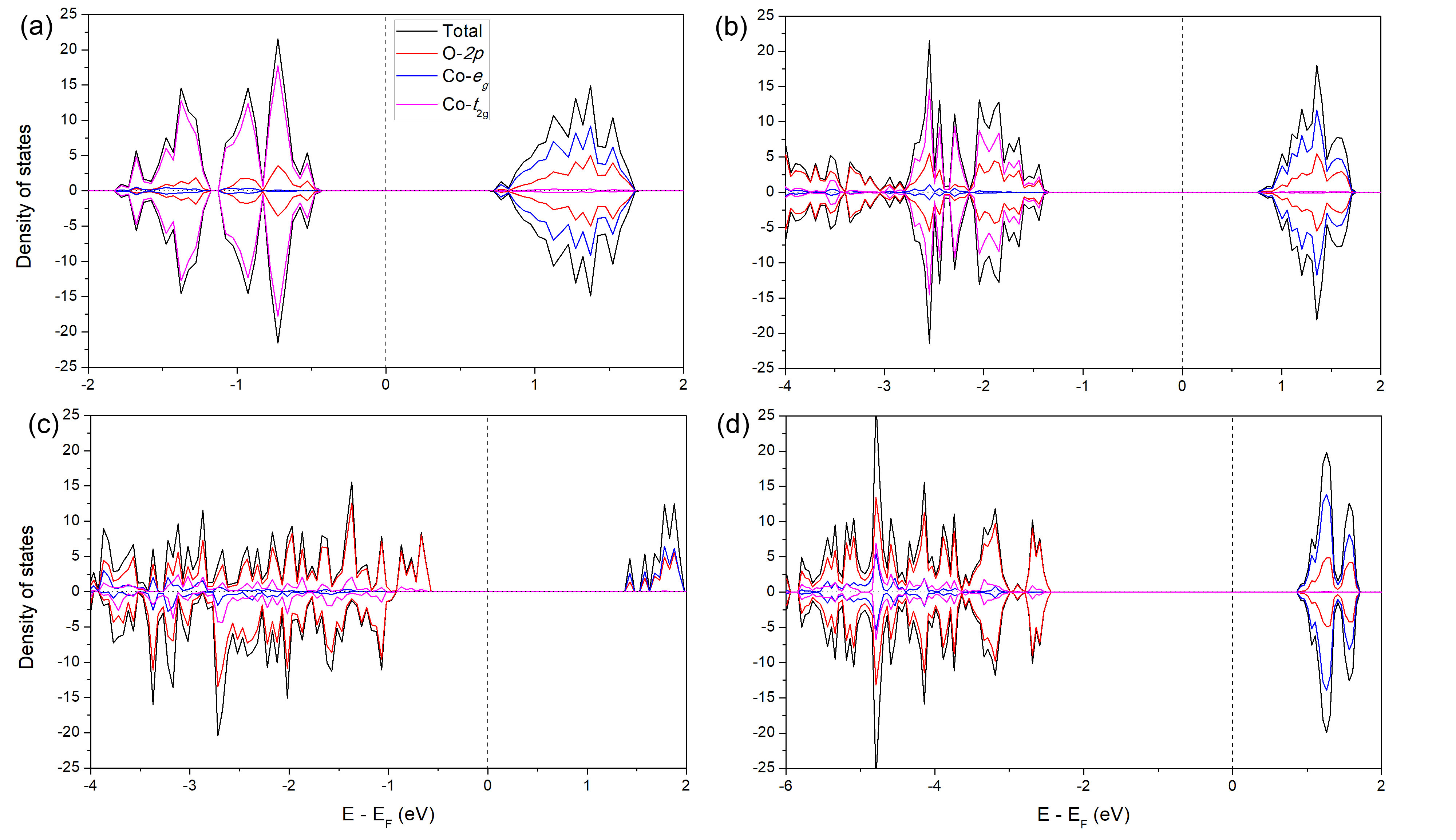}
 \caption{DFT density of states of LiCoO$_2$ for (a) PBE, PBE+U with U = (b) 3 eV, (c) 10 eV, and (d) 10 eV with constraints for zero magnetization.}

 \label{fig:DFT_DOS}
\end{figure}
In Fig.~\ref{fig:DFT_DOS}(a) and (b), the strong $p$-$d$ hybridization can be observed in the valence band edge in the PDOS, and the band gap is wider for PBE+U due to the presence of on-site Coulomb repulsion from the Hubbard U.
Note that the value of U = 3.0 eV in Fig.~\ref{fig:DFT_DOS}(b) is taken from the effective U reported in a previous PBE+U study for LiCoO$_2$.~\cite{dixit16}
As Hubbard U increases to a large value beyond 10 eV, PBE+U produces an unbalanced density distribution between the two spin channels, yielding an anti-ferromagnetic state as shown in Fig.~\ref{fig:DFT_DOS}(c).
In order to obtain a low-spin state, we repeated the calculation using PBE+U with U=10 eV while constraining the total magnetization to be zero.  This yields similar $p$-$d$ hybridization to the other PBE+U (see Fig~\ref{fig:DFT_DOS}(d)), however, its total energy of -4840.511 eV per formula unit (f.u.) of LiCoO$_2$ is energetically unfavorable as compared to -4840.531 eV/f.u. total energy in the unconstrained system.
The Hubbard occupancies of $t_{2g}$ and $e_{g}$ states with U = 2 - 10 eV are 5.93 - 5.98 and 1.75 - 1.59, respectively, which is consistent with previous DFT+U and DFT+DMFT results of 5.8 ($t_{2g}$) and 1.6 ($e_{g}$).~\cite{issacs2020}
We additionally computed L\"{o}wdin populations for comparison to these Hubbard occupations.
Computed populations for the $t_{2g}$ and $e_{g}$ states are 5.88 - 5.97 and 1.48 - 1.28 for U = 0 - 10 eV, respectively (see Supplemental Information for greater detail).~\cite{SI}
From this we see that the main effect of introducing the on-site Coulomb repulsion is to reduce the Co $e_g$ occupation toward the idealized ionic picture.
The fact that such a large U is required to partially deplete the $e_g$ states suggests that the energy separation between the t$_{2g}$ and e$_g$ states near the Fermi level 
in LiCoO$_2$ may be difficult to describe within DFT+U.

\begin{figure}[t]
 \includegraphics[width=6 in]{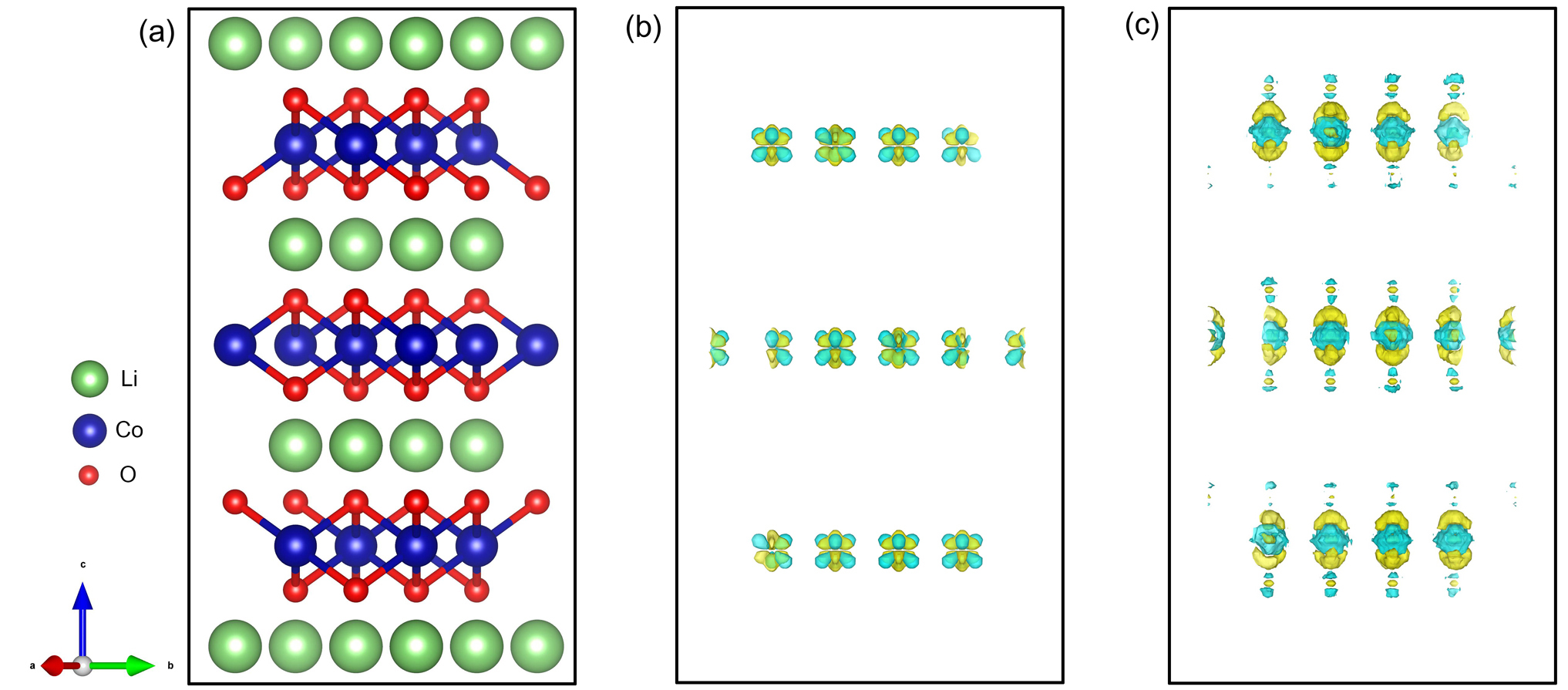}
 \caption{(a) 12 formula unit supercell of LiCoO$_2$ and its isosurface of charge density difference $\Delta \rho$ for (b) $\rho(\text{PBE+U}_{opt})-\rho(\text{PBE})$ and (c) $\rho(\text{DMC})-\rho(\text{PBE+U}_{opt})$.  }
 \label{fig:Density}
\end{figure}

Figure~\ref{fig:Density} represents isosurfaces of charge density difference of LiCoO$_2$ between PBE, PBE+U$_{opt}$, and DMC. Note that density isosurface levels in this figure are set to the same value of 6x10$^{-5}$ $|e|\slash \text{\AA}^{3}$. 
As the on-site Coulomb interaction is taken into account by adding a Hubbard U repulsive term to the Kohn-Sham Hamiltonian, we observe a change in the distribution of the 3$d$ electrons on the Co atom in LiCoO$_2$.
In Fig~\ref{fig:Density}(b), the density redistribution near the Co atoms induced by the Hubbard repulsion shows charge accumulation and depletion simultaneously on the direction of triplet $t_{2g}$ ($d_{xy}$, $d_{xz}$, and $d_{yz}$) state.
On the other hand, the DMC density shows a large charge accumulation along both the positive and negative $z$ direction near the Co atoms which is induced by a collective $p$-$d$ hybridization with the nearby O atoms.
Figure~\ref{fig:Density}(c) shows that there is significant difference in the charge density on O site between DMC and PBE+U; DMC gives more strongly correlated electron density distributions between Co and its nearest neighbor O atoms.
In conclusion, DMC charge density shows strong $d_{z^{2}}-{p}$ orbital coupling, leading to predict less occupation on $e_{g}$ state in DMC than PBE+U. 
\begin{figure}[t]
 \includegraphics[width=6 in]{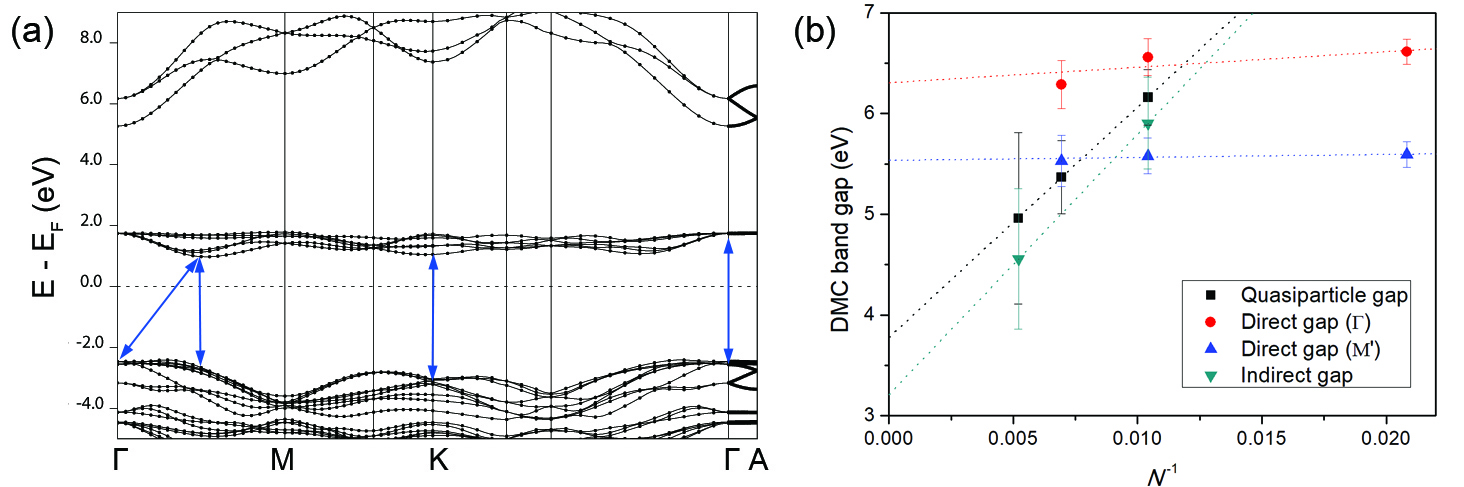}
 \caption{(a) PBE+U band structure of bulk LiCoO$_2$. Blue lines indicate three direct gaps and one indirect gap in the band structure. (b) DMC band gaps of LiCoO$_2$ as function of 1/$N$, where $N$ represents total number of atoms in the supercell. Dotted lines indicate the simple linear regression fit. 
 }
 \label{fig:SD_band}
\end{figure}

To study the excited state for LiCoO$_2$, we first computed the LiCoO$_2$ band structure using PBE+U$_{opt}$ in order to determine the locations of the valence band maximum (VBM) and conduction band minimum (CBM) for subsequent DMC calculations of the gap.
In Fig.~\ref{fig:SD_band}(a), LiCoO$_2$ exhibits insulating band structure with wide band gaps are in PBE+U because of the effect of Hubbard U repulsion on the valence band.
We calculated the DMC quasiparticle and optical gaps at selected k-points for the direct and indirect gap because conduction bands around Fermi level.
We computed quasiparticle gap by using the equation $E_{qp} = E(N+1) + E(N-1) - 2E(N)$ where $E(N)$ represents total energy for $N$ electrons system.
DMC excitonic gap can be estimated by computing energy difference between the ground state and excited state energy $E_{ex}(k)-E_{g}(k)$ where $E_{ex}$ represents the particle-hole excitation energy obtained by promoting a particle to the excited state at the specific k-point.
We here choose two different direct gap ($\Gamma$ (0.00,0.00,0.00) and M' (0.25, 0.00, 0.00)) and an indirect gap at $\Gamma$-M'. (See Fig.~\ref{fig:SD_band}(a))

We observed that while a large value of U provides an accurate description of the ground state, excited state energies computed within DMC based on particle-hole type promotions between the DFT+U band edges were largely overestimated.  
In contrast, we found that excitations calculated in DMC with plain PBE wavefunctions provides a better description of the excitation energies, indicating that the single particle states at either the valence or conduction band edges may be poorly described in DFT+U.
Detailed results for DMC band gaps as function of U in wavefunction can be found in Supplemental Information.~\cite{SI}
As seen in Fig.~\ref{fig:SD_band}(b), we estimated DMC optical gaps at the thermodynamic limit by extrapolating DMC gaps to the bulk limit of $N = \infty$ where $N$ is the total number of atoms in a supercell.
\begin{table*}[t]
\centering
\caption{Calculated DMC and DFT+U (U = 0, 3.0 eV, and U$_{opt}$) direct and indirect gap for bulk LiCoO$_2$ at the selected k-points. Energy units are given as eV.
}
\label{tab:bandgap}
\begin{tabular}{c|ccc|ccc|c|c}
\hline\hline
                   &  \multicolumn{3}{c|}{LDA+U (eV)} & \multicolumn{3}{c|}{PBE+U (eV)} & \multirow{2}{*}{DMC} & \multirow{2}{*}{Exp.} \\ \cline{2-7}
                   &  0 (LDA)  & 3.0  & 13.8  & 0 (PBE)  & 3.0  & 11.0  & & \\ \hline 
direct gap(M') & 0.92 & 1.92 & 3.55 & 1.24 & 2.28 & 3.64 & 5.5(3) & \\
direct gap($\Gamma$) & 2.05 & 3.02 &  4.15 & 2.26 & 3.31 & 4.19 & 6.3(3) & 2.1 $\sim$ 2.7$^{1}$ \\
indirect gap($\Gamma$-M')  & 1.11 & 2.11  &  3.34 & 1.42 & 2.47  &  3.43 & 3.2(10)   &    \\  \hline
quasiparticle gap &   &  &  &  &  &  & 3.7(9)  & \\ \hline \hline
\end{tabular}
\begin{flushleft}
1. Ref.~\cite{elp91,rosolen01,ghosh06,rao09,ensling10}.\\
\end{flushleft}
\end{table*}

Table~\ref{tab:bandgap} summarize computed DFT optical gaps and DMC ones from PBE wavefunction for the selected k-points.  
Underestimated band gap in LDA or PBE compared to the experimental result is not surprising issue as many DFT studies already have claimed.~\cite{shin17}
When the Hubbard correction U is introduced in DFT scheme, we see large gap opening in both PBE+U and LDA+U results and LDA+U and PBE+U band gap energy scale become similar with the experimental value on U = 3.0 eV.
On the other hand, at U$_{opt}$, it is found that both PBE+U and LDA+U shows significantly larger band gaps than the experimental result.
From no significant changes on Hubbard occupation for $t_{2g}$ and $e_{g}$ state with different values of U up to 10 eV, we conclude that the existence of Hubbard U on LiCoO$_2$ significantly open the band gap, but it doesn't give much effect on 3$d$ orbital distribution in DFT+U scheme as shown in the Supplemental Information.
In DMC results, the estimated indirect gap(3.2(10) eV) is statistically consistent with the experimental results of 2.1 - 2.7 eV.~\cite{elp91,rosolen01,ghosh06,rao09,ensling10}
However, we found that DMC direct gaps at both of $\Gamma$ and M' point are
estimated to be much larger than the indirect gap, leading us to suspect that the nodal surface from DFT is not optimized to describe the excited state properties of LiCoO$_2$.
We have confirmed that addition of variational parameter U is enough to provide a highly accurate ground state energy within single-determinant PBE+U trial wavefunction via comparison of the cohesive energy to corresponding experimental result, however, this Hubbard U is not suitable as the variational parameter to optimize nodal surface of the excited state since increasing value of U monotonically increases optical and quasiparticle gaps (see Supplemental Information).
In order to understand the large discrepancy between the DMC direct gap and the experimental results, re-visiting estimation of direct gap for LiCoO$_{2}$ with trial wavefunctions beyond the single determinant PBE+U level is highly desirable.

\subsection{Analysis based on selected-CI wavefunctions}

In single reference, PBE+U orbitals provided the excellent nodal surface for the ground state, but it is not guaranteed PBE+U wavefunction is optimal to minimize FN error for the excited state as monotonically increasing DMC band gap was shown with increasing value of U in the PBE+U trial wavefunction.
In order to investigate the dependency of the number of determinants on FN error for the excited state, we consider multi determinants wavefunction based on gaussian basis-set. 
In this study, a set of NOs are generated for both single and the multi references in a trial wavefunction to compare the quality of nodal surface with corresponding DFT orbitals.

\subsubsection{CIPSI wavefunction convergence \label{sec:CIPSI_conv} }
\FloatBarrier

To generate trial wavefunctions for multideterminants, we employ the sCI method within its CIPSI implementation. For systems of relatively compact size, where the variational energy of both ground and excited states—with or without the renormalized second-order energy correction (rPT2)—can be effectively converged by adjusting the number of determinants, one can achieve an accurate representation of the difference in energy between these states. In such cases, employing state-averaged density matrices obtained from these multideterminant expansions facilitates the derivation of natural orbitals (NOs), significantly reducing computational costs while enhancing the quality of the original orbitals. For the simultaneous treatment of ground and excited states, setting the selection algorithms to achieve a balanced rPT2 value between both states ensures a robust error compensation mechanism, maintaining accuracy for both states\cite{Filippi2022}.

However, when dealing with larger system sizes, such as those explored in our present study, achieving convergence with respect to the number of determinants becomes challenging, as evidenced in the Supplementary Information (SI) Fig.~5.
For this reason, we performed several sCI calculations initialized with different orbital sets.  
The initial orbital sets were obtained from PBE+U (with U=0, 3, 5, 7, 11 eV) and we denote sCI performed with these as sCI/PBE+U=X.
The generation of NOs from unconverged multideterminant expansions cannot be assumed as complete, considering that the inclusion of missing determinants could potentially lead to significant improvements in the set of orbitals. 
Consequently, our approach to generating NOs is iterative: we initiate a multideterminant expansion with a few million Slater determinants, derive the NOs from state-averaged density matrix, and repeat this process until convergence of the E+rPT2 energy is attained. However, implementing this procedure with an equal weighting on both ground and excited states does not maintain error compensation, as the ground state tends to converge notably faster than the excited state, as indicated in SI Fig.~6. 

\begin{figure}[t]
 \includegraphics[width= 5in]{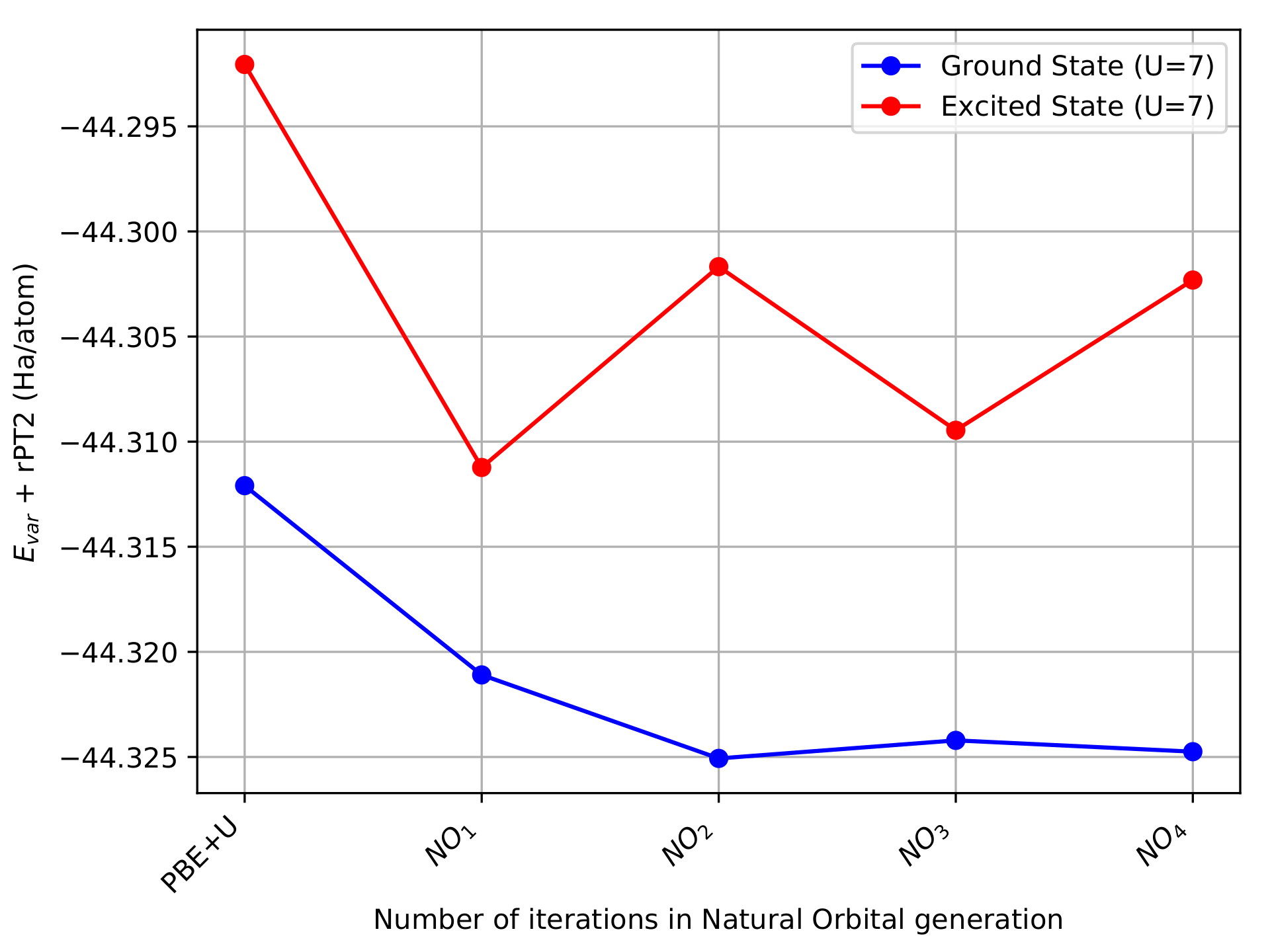}
 \caption{Variational Energy + rPT2 of the ground and excited state as a function of the number of iterations used to generate natural orbitals for  sCI/PBE+U=7.}
 \label{fig:EPT2-NO-U7}
\end{figure}

\begin{figure}[htbp]
  \centering
   \begin{subfigure}[b]{0.8\textwidth}
    \centering
    \includegraphics[width=\textwidth]{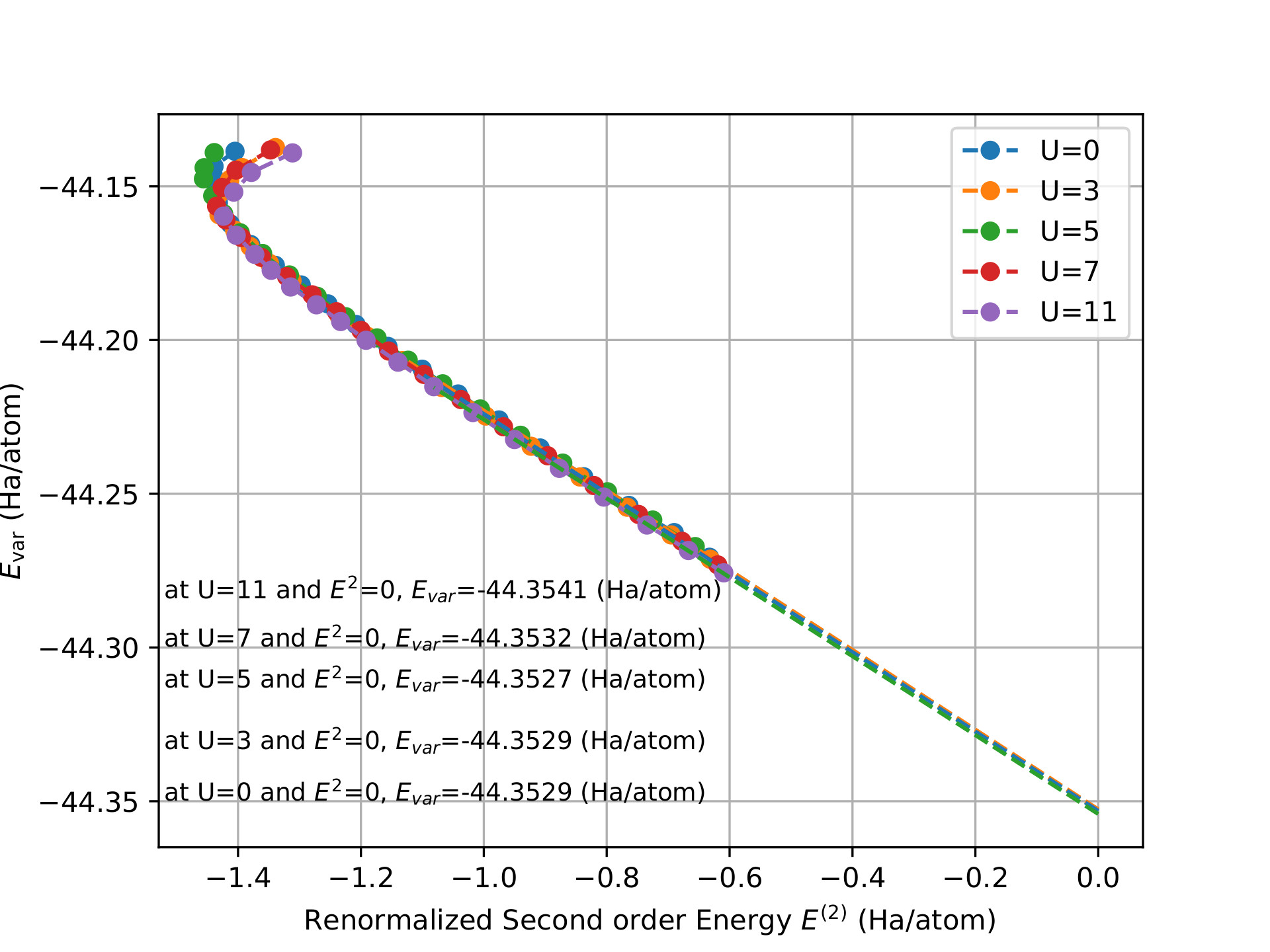}
    \caption{Ground State}
    \label{fig:EvarVSrPT2b}
  \end{subfigure}
  \hfill
  \begin{subfigure}[b]{0.8\textwidth}
    \centering
    \includegraphics[width=\textwidth]{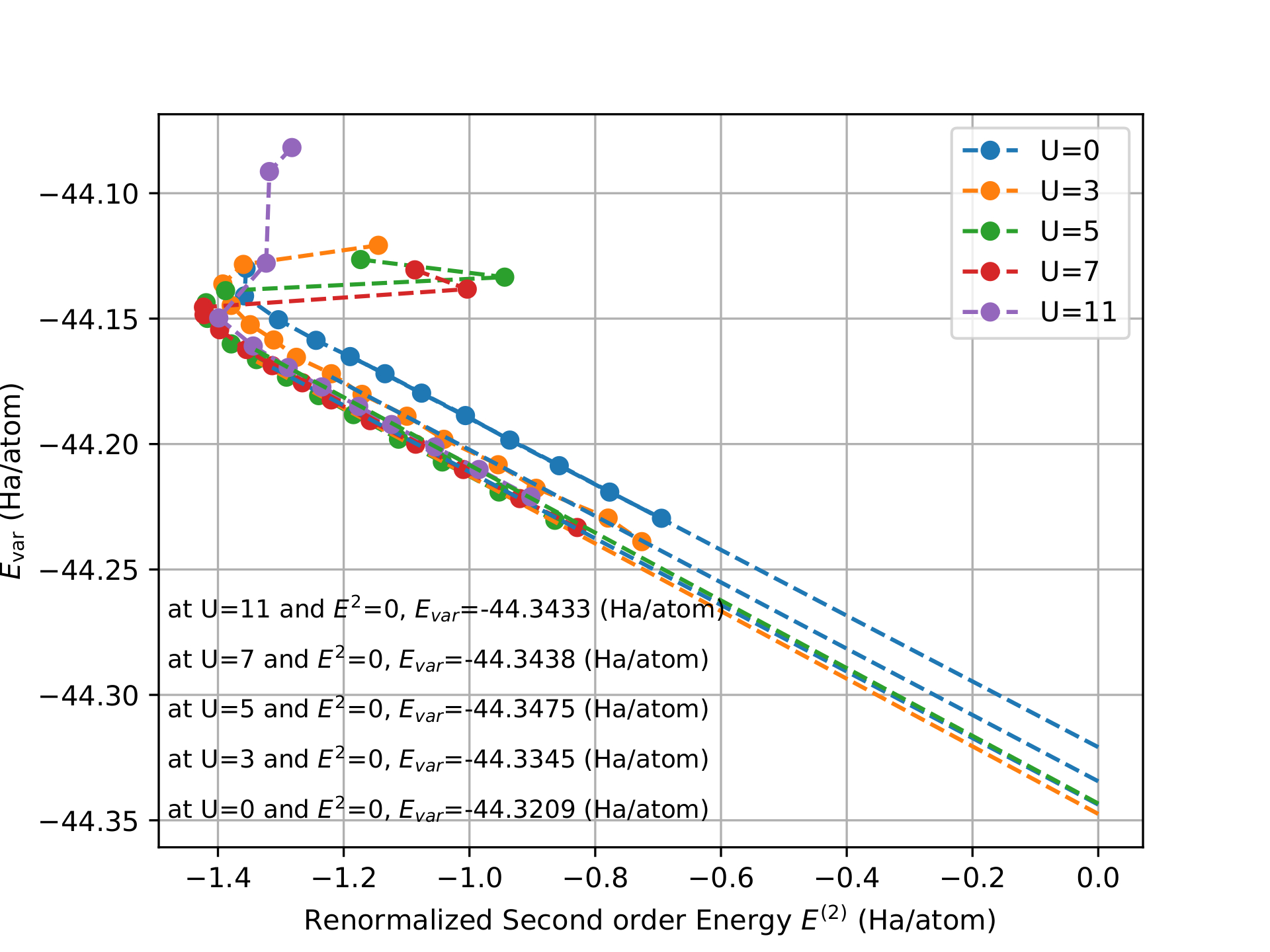}
    \caption{Excited State}
    \label{fig:varVSrPT2a}
  \end{subfigure}
   \caption{Variational energy $E_{var}$ for the ground state (top) and excited state (bottom) as a function of the renormalized second-order energy $E_{var}$+ rPT2 ($E^{(2)}$) for the largest wavefunctions in the converged natural orbitals for sCI/PBE+U for various values of U. A linear fit  of the last eight points is also reported with the extrapolated values of the variational energy at rPT2=0.}
  \label{fig:EvarVSrPT2} %\textcolor{red}{Can we make this figure bigger?}
\end{figure}

This discrepancy necessitates extensive multideterminant expansions to balance the value of rPT2 for both states. Consequently, we opt for independent convergence runs for the ground and excited states, as can be seen in Fig.~\ref{fig:EPT2-NO-U7} for sCI/PBE+U=7 and for all other wavefunctions in the SI Fig.~7.
This choice is further motivated by the exponential scaling of the sCI method, where efficiency is gained by running two smaller, independent computations for a larger number of determinants instead of a single extensive one. The effect of the NO iterative scheme on both ground and excited states is observable; in both cases, the total energy and quality of the orbitals improve even after a single iteration of the NOs over the original Kohn-Sham (KS) single-particle orbitals. While the ground state, across various KS states, converges almost to the same energy, the excited state exhibits significantly more differences depending on the starting wavefunction (value of U). However, after 5 iterations, the last set of NOs seems to have converged for both ground and excited states, giving a spread of 2 mHa/atom and 4 mHa/atom among final wavefunctions for the ground and excited states, respectively (excluding sCI/PBE+U=0 for the excited state due to poor quality). Adding more iterations over the NOs would not significantly change the variational energy beyond a few mHa/atom. 
To select the optimal orbitals describing our system, we extrapolate the variational energy to an rPT2 value of 0 (Fig.~\ref{fig:EvarVSrPT2}). It's evident that after the convergence of the NOs, the ground state of LiCoO$_2$ converges to sub-milliHartree values for all U values. In contrast, the excited state displays greater dependence on the initial conditions, with an increase in the value of U leading to the lowest energy. The largest discrepancies (over 10 mHa/atom) are observed between sCI/PBE+U=0/3 eV and the others. The difference between sCI/PBE+U=7 eV and sCI/PBE+U=11 eV is only 0.5 mHa/atom, rendering both systems relatively indistinguishable. Plotting the energy against the CI variance (SI Fig.~8)
further confirms that the ground state is converged to the same wavefunction regardless of the starting KS orbitals, while the excited states exhibit more pronounced differences. Although sCI/PBE+U=7 eV and sCI/PBE+U=11 eV are quite similar, the expansion for U=7 eV appears slightly more compact. Therefore, we select sCI/PBE+U=7 as the best trial wavefunction for the remainder of the analysis.  Below, results for this wavefunction appear as sCI-MD (multi-determinant) while results obtained from the single reference state constructed from its natural orbitals appear as sCI-SD (single determinant).

\FloatBarrier

\subsubsection{Nature of the $\Gamma-\Gamma$ excitation in LiCoO$_2$}
\FloatBarrier

The accurate multideterminant wavefunction we have obtained via sCI allows us to characterize both the many body ground state of LiCoO$_2$ and its lowest lying excitation at the $\Gamma$-point.  Atomic L\"{o}wdin charges for the sCI-MD (i.e. sCI/PBE+U=7) ground and excited states are collected in Table \ref{tab:lowdin_cipsi}.  The ground state of LiCoO$_2$ is characterized by a nearly full Co-$t_{2g}$ sub-manifold along with low Co-$e_g$ filling.  Oxygen is under-compensated, relating to the presence of ligand holes in the material.  The nearest ionic-like filling is Co-$t_{2g}^6e_g^1$ O-$p^5$.  The lowest lying excitation at the $\Gamma$-point is primarily a Mott-like $d$ to $d$ transition involving a promotion from Co-$t_{2g}$ to Co-$e_g$ states, similar to what is shown in the ionic-like picture of Fig. \ref{fig:LiCoO2}, but in the presence of a partially filled Co-$e_g$ band.  The excitation is accompanied by a slight reduction in the ligand hole, as shown by the increasing O-$p$ occupation, corresponding to about 7\% of the redistributed weight due to the excitation.

\begin{figure*}[t]
\centering
\includegraphics[width=3 in]{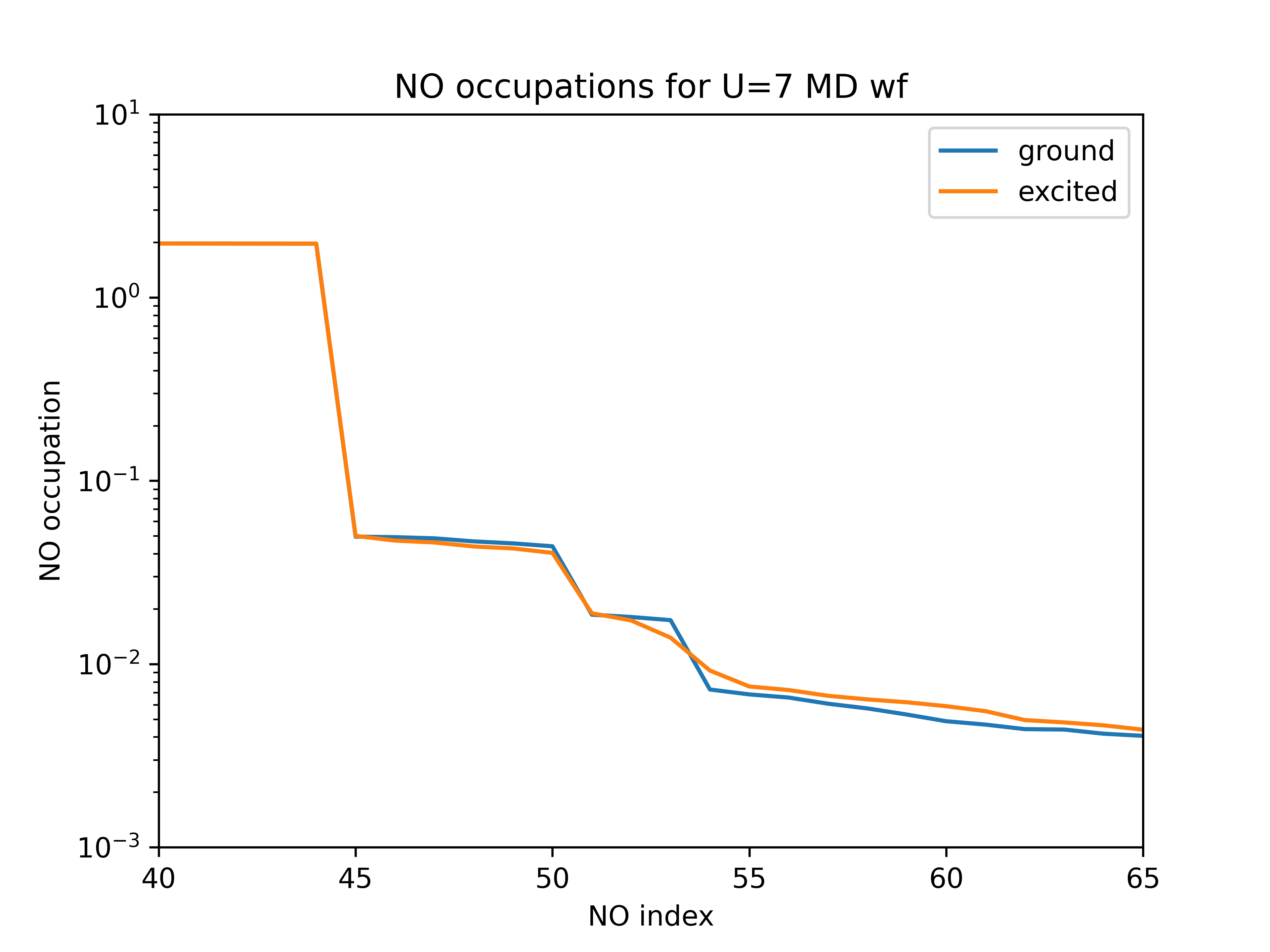}
\includegraphics[width=3 in]{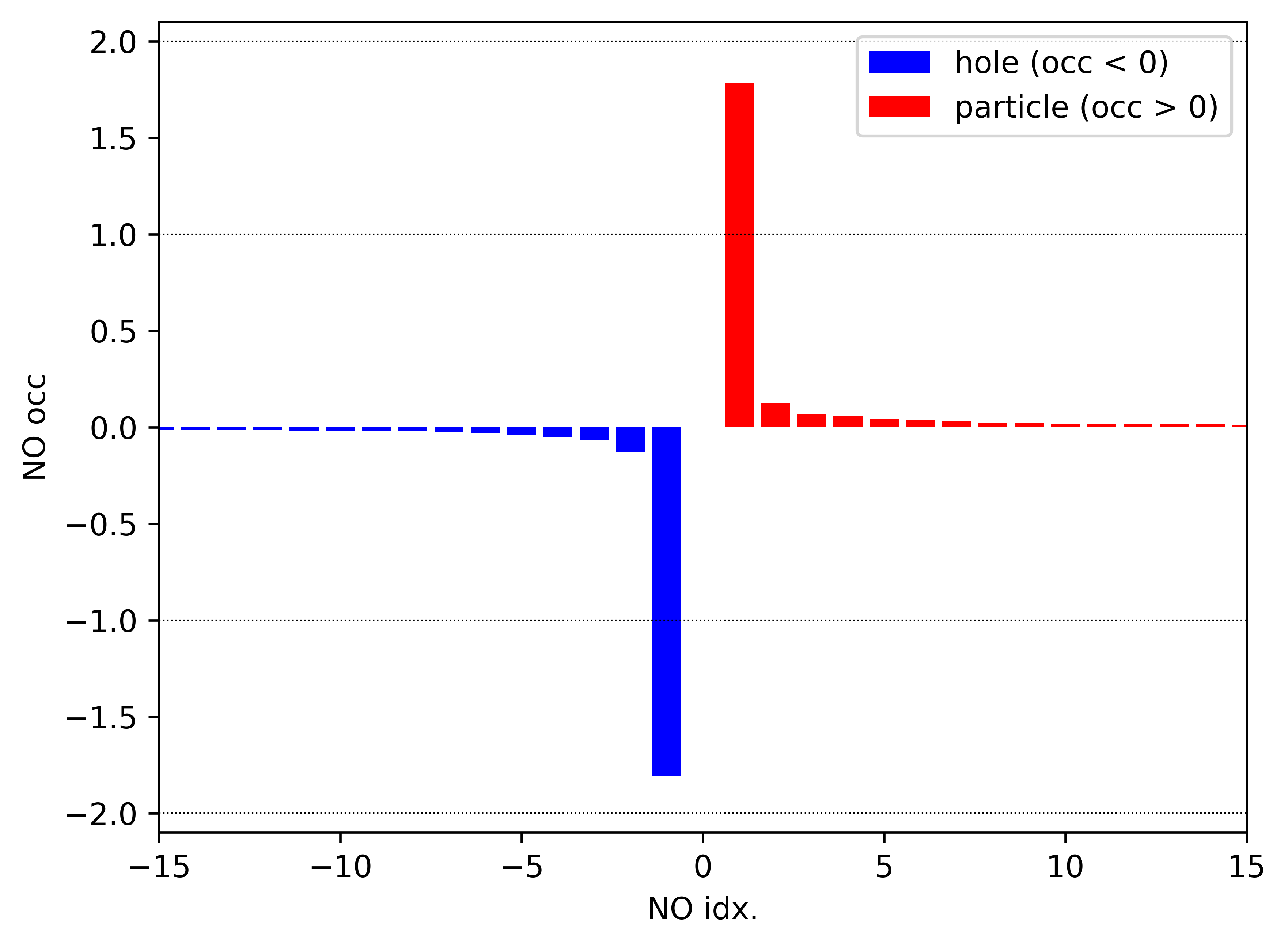}
\caption{(a) Natural occupation numbers for sCI-MD ground and $\Gamma$-point excited states for LiCoO$_2$ in descending magnitude, (b) Particle (red) and hole (blue) occupation numbers for the $\Gamma$-point excitation obtained from the density matrix difference between sCI-MD excited and ground states. %\textcolor{red}{please lable (a) and (b) for figurez.}
}
\label{fig:MD-U7-diff-NO-bar}
\end{figure*}

In order to probe the degree of multireference character in these states, we constructed natural orbitals and natural occupation numbers for each state by diagonalizing their respective 1-body reduced density matrices.  The resulting occupation numbers, ordered in descending magnitude, are shown in Fig. \ref{fig:MD-U7-diff-NO-bar}a.  In the limit of vanishing multireference character, each state would be fully described by a single Slater determinant.  Our unit cell of LiCoO$_2$ contains three formula units (three layers) and 90 electrons and so a single reference state would appear as 45 doubly occupied natural orbitals.  As can be seen in Fig. \ref{fig:MD-U7-diff-NO-bar}a, both the ground and excited states have a dominant single reference component of this type, but with significant multireference contributions from many other states with rapidly decreasing weight.  A long, smooth evanescent tail of this type is a hallmark of dynamic correlation that is present in all materials.  In addition to this tail (beginning roughly with natural orbital 55 and higher), we also see two distinct plateaus of higher weight involving 1-2 states per layer which may be  associated with static correlation effects.  

The impact of correlation on the physical states can be assessed by removing the contribution from the dominant single reference state (sCI-SD) from the atomic L\"{o}wdin charges, as also contained in Table \ref{tab:lowdin_cipsi} ($\Delta_{\textrm{sCI-SD}}^{\textrm{sCI-MD}}$).  In both the ground and excited state, multireference correlation promotes negative charge transfer (O-$p\rightarrow$Co-$e_g$) in LiCoO$_2$ which increases the magnitude of the ligand hole charge.  For the excited state, an even larger effect is seen in the multireference enhancement of the t$_{2g}$ to $e_g$ transition in the excitation.  

Since the excitation occurs within the partially filled $d$-manifold of Co in Mott fashion, it is expedient to clarify whether the excitation is collective in nature or of particle-hole type.  We evaluate this question by analyzing the eigenstates of the difference density matrix\cite{HeadGordon1995} of the many body states.  The resulting difference occupation numbers are shown in Fig. \ref{fig:MD-U7-diff-NO-bar}b.  On this plot, particle-hole pairs appear as bars of equal height and opposite sign.  From this we see that the $\Gamma$-point excitation in LiCoO$_2$ is primarily of particle-hole type as a single pair accounts for about 85\% of the weight.  The collective contribution (15\%) is quite substantial, however, which suggests simple mean-field theories may struggle to faithfully represent the spectral properties of LiCoO$_2$.

\begin{table*}[t]
\centering
\begin{tabular}{c|ccc}
\hline\hline
                 &   Co-$t_{2g}$  &  Co-$e_g$ & O-$p$  \\   \hline
  Ground   sCI-MD    &   5.941  & 1.108  &  9.669          \\
  Excited  sCI-MD    &   5.478  & 1.524  &  9.702          \\
  $\Delta_{\textrm{Ground}}^{\textrm{Excited}}$  sCI-MD    &  -0.462  & 0.416  &  0.033      \\
  Ground   $\Delta_{\textrm{sCI-SD}}^{\textrm{sCI-MD}}$ &   0.006  & 0.036  & -0.044          \\
  Excited  $\Delta_{\textrm{sCI-SD}}^{\textrm{sCI-MD}}$ &  -0.093  & 0.112  & -0.031         \\ \hline \hline
\end{tabular}
\caption{L\"{o}wdin populations of Co-$t_{2g}$, Co-$e_{g}$, and O-$p$ orbitals found by CIPSI for LiCoO$_2$ ground and $\Gamma$-point excited states.  sCI-MD refers to fully multireference CIPSI states (from sCI/PBE+U=7 calculations), while sCI-SD refers to the single determinant state constructed from the sCI-MD natural orbitals.  Differences in atomic population between excited/ground states ($\Delta_{\textrm{Ground}}^{\textrm{Excited}}$) and multideterminant/single-determinant ($\Delta_{\textrm{sCI-SD}}^{\textrm{sCI-MD}}$) states are also provided.}
\label{tab:lowdin_cipsi}
\end{table*}

\FloatBarrier

\subsubsection{Assessment of Kohn-Sham orbital quality}
\FloatBarrier

Another metric by which we can assess the quality of the PBE orbitals is to determine how well they can describe the excitation as compared to the sCI-MD excited state.

\begin{figure}[t]
\includegraphics[width=6 in]{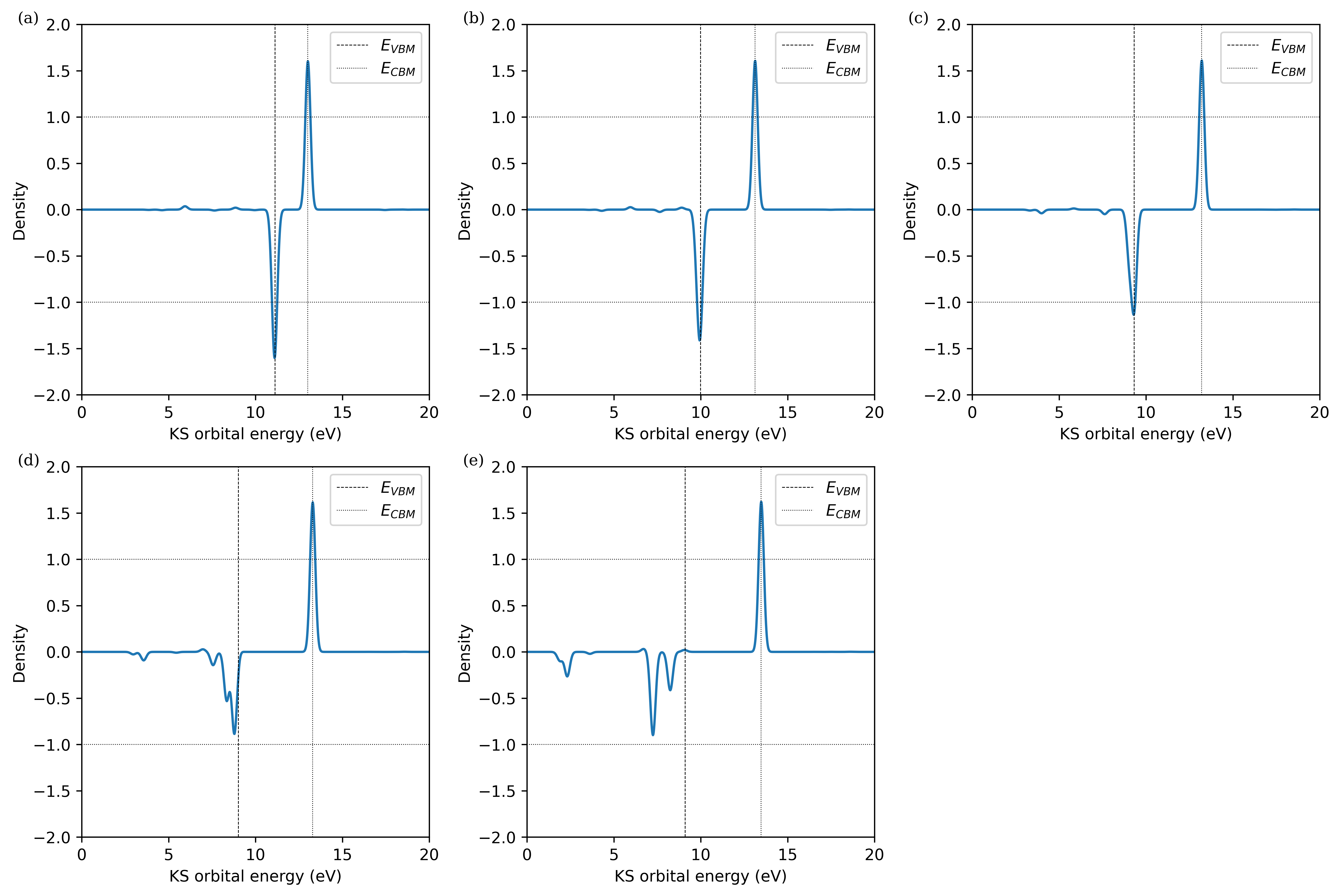}
\caption{DOS-like single particle occupancy difference between sCI-MD excited and ground states projected onto PBE+U orbitals with U=(a) 0, (b) 3, (c) 5, (d) 7, (e) 11 eV and organized according to the respective Kohn-Sham (KS) eigenvalues.  In each case, the PBE+U VBM and CBM are shown as vertical dashed lines.  A DFT functional with an ideal representation of the sCI-MD particle-hole-like excitation would appear as two isolated negative/positive peaks at the KS VBM/CBM.}
\label{fig:MD-U7-pbe-dos}
\end{figure}

By taking the difference between ground and excited state sCI-MD wavefunctions, we can get the single particle density matrix difference corresponding to this particle-hole excitation.
This density matrix difference can be projected onto a set of KS orbitals to determine whether the particle and hole states will be accurately represented by the CBM and VBM obtained within PBE+U.  In cases where PBE+U accurately captures the dominant particle-hole excitation found in sCI-MD, the difference in projected single particle occupancies should appear as a single positive peak at the CBM and a single negative peak at the VBM.
Fig \ref{fig:MD-U7-pbe-dos} shows the result of projecting the sCI-MD ground/excited state difference onto the PBE+U orbitals obtained using several values of U.  In the Figure, the projected occupancies have been arranged based on the KS eigenvalue similar to standard DFT DOS plots with the location of the PBE+U VBM and CBM appearing as vertical dashed lines.

For all values of U, the sCI-MD particle state (positive peak) is well-described by the PBE+U CBM; therefore, all of these sets of KS orbitals provide a good description of the final state for the $\Gamma-\Gamma$ excitation in LiCoO$_2$.
In contrast to the isolated particle state at the CBM for all values of U, the location of the hole state within the PBE+U band structure varies dramatically with changes in U.
For U=0, the sCI-MD hole state (negative peak) is almost entirely contained at the PBE VBM, but as U increases, the hole state moves deeper into the valence band and becomes spread across several of the KS orbitals in the PBE+U valence band.
For small positive values of U (3-5 eV), the hole is still near the PBE+U VBM with only a small broadening, but by U=7 eV a significant portion of the hole that has moved below the band edge.
At U=11 eV there is virtually no hole density at the PBE+U VBM, so the hole state will be completely misrepresented by a single particle-hole excitation in these orbitals.

\begin{figure}[t]
\includegraphics[width=6 in]{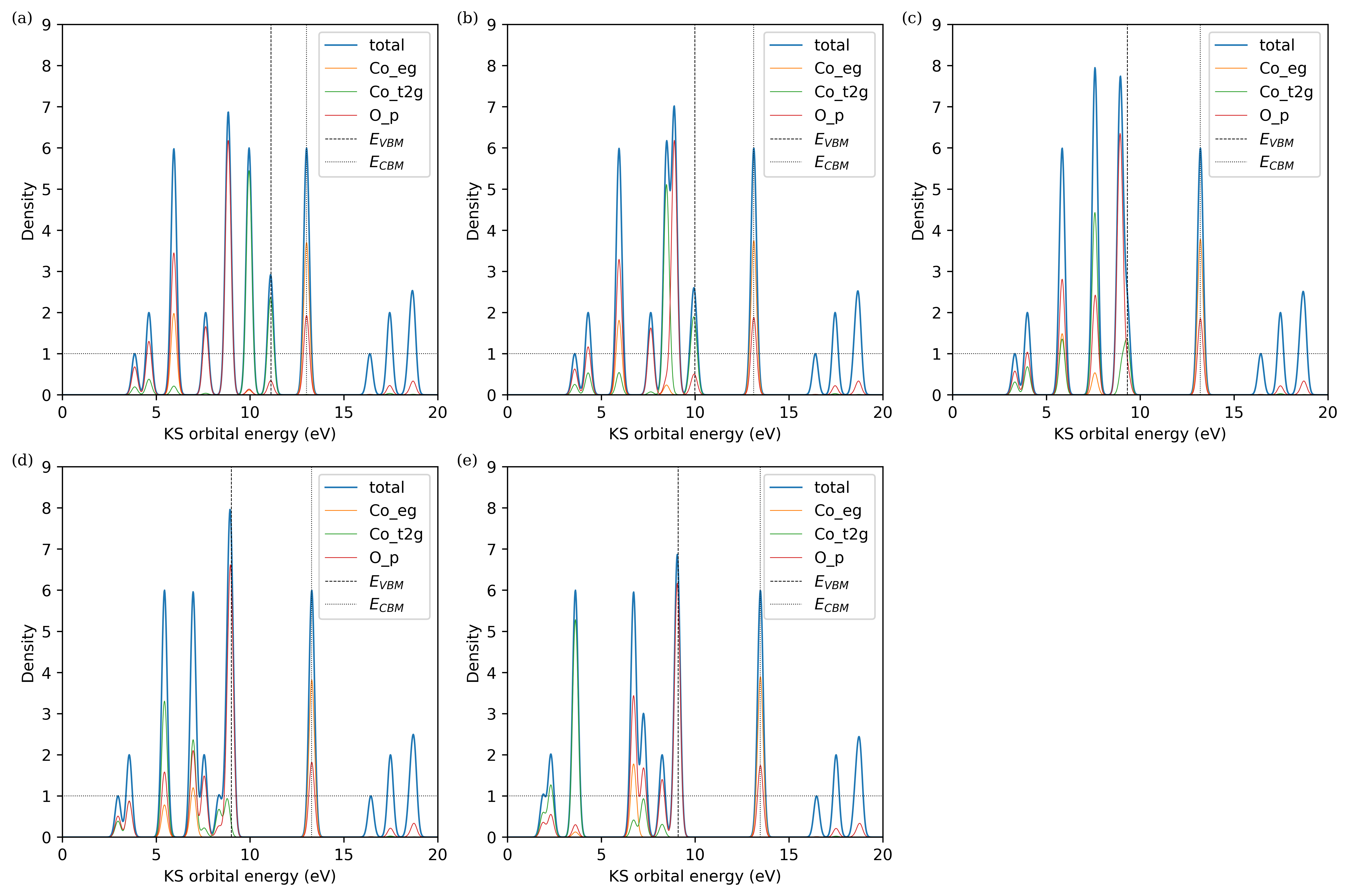}
\caption{PBE+U $\Gamma$-point PDOS showing contributions from cobalt $d$ and oxygen $p$ orbitals for U=(a) 0, (b) 3, (c) 5, (d) 7, (e) 11 eV.  In each case, the PBE+U VBM and CBM are shown as vertical dashed lines.}
\label{fig:pbe-lcao-dos}
\end{figure}

Fig \ref{fig:pbe-lcao-dos} shows the PBE+U $\Gamma$-point PDOS, highlighting from the cobalt $d$ and oxygen $p$ atomic orbitals. As expected, the nature of the CBM does not change significantly as U increases, retaining a dominant Co-$e_g$ character with some O-$p$ hybridization.  However, the character of the PBE+U VBM changes drastically with varying U.  There is a large Co-$t_{2g}$ contribution at the VBM for U=0.  As U increases to 3 eV, the Co-$t_{2g}$ state begins to move slightly deeper into the valence band, and at U=11 eV it is far below the band edge and hybridized with multiple other states.  At U=11, the VBM consists almost entirely of oxygen $p$ character, which misrepresents the particle-hole excitation as charge transfer type ($p$-$d$), rather than Mott type ($d$-$d$).  

Interestingly, the sCI-MD hole state appears as a well isolated single state at the VBM of plain PBE.  Thus the simplest of the functionals, in terms of on-site correlation, best describes the lowest energy gamma excitation in LiCoO$_2$.  This situation contrasts to the best description of the ground state as found by DMC, which we explain in detail next.

\FloatBarrier

\subsubsection{Validation of high optimal U for the DMC ground state}
\FloatBarrier

In the previous section, we showed that the PBE+U CBM/VBM obtained at larger U values do not accurately describe the true particle-hole excitation as found by sCI.
In contrast to this, we see that the \textit{total} occupations of Co $t_{2g}$, $e_g$, and O $p$ atomic orbitals found with PBE+U more closely match those from sCI at higher U values.
As shown in Table~\ref{tab:lowdin_pbeu_vs_cipsi}, L\"{o}wdin population (occupation) in each of these atomic states gets closer to the sCI-MD occupations as U increases.
Compared to the sCI-MD wavefunction, the Co-$t_{2g}$ and O-$p$ occupations are underestimated by PBE+U at all values of U, and the Co-$e_{g}$ is overestimated.
Comparison of sCI-MD occupations with a previous DFT+DMFT study reveals a similar findings: a minimal disparity in $t_{g}$ occupation, but a significant overestimation in the Co-$e_{g}$ occupancy in DFT+DMFT. For instance, at U = 6 eV,\cite{issacs2020} the $e_{g}$ occupation is approximately 1.4, similar to our PBE+U $e_g$ occupation at similar values of U.

The sCI-MD atomic occupations are very close to the idealized complete filling of the Co-$t_{2g}$ subshell.
Interestingly, the observed increase in O$_{p}$ occupation as U increases or when comparing PBE+U with sCI-MD aligns quantitatively with the decrease in $e_{g}$ occupation. 
A plausible interpretation of these findings is that larger U values appropriately penalize $e_{g}$ occupation, resulting in a more complete t$_{2g}$ subshell, and a more moderate Co-O ligand hole.

This trend also correlates with the DMC total energies obtained when using PBE+U orbitals to construct single-determinant trial functions. There, the DMC energies improve with increasing U (variationally lower), as shown in Fig. \ref{fig:DMC_U}(a).
These results suggest that DMC and sCI treatments are in full agreement in terms of treatment of Co $t_{2g}$, $e_g$, and O $p$ populations; however, because DMC also relies on DFT for a first-order accurate description of the excitation, it will be biased by inaccuracies in the DFT.  This is consistent with what we observe at higher U for DMC: large overestimation of excitation energies, consistent with the poor quality of the PBE+U VBM state.

Taken together, these results show that DFT cannot be relied on to simultaneously describe both the ground state and the low-lying excitations of LiCoO$_2$.
This underscores the importance of moving beyond DFT for an accurate description of trial states for excitations within real-space QMC methods.

\begin{table*}[t]
\centering
\begin{tabular}{c|ccc|ccc}
\hline\hline
         & Co-$t_{2g}$ & Co-$e_g$ & O-$p$   &  $\Delta(t_{2g})$ &  $\Delta(e_g)$ &  $\Delta(p)$  \\ \hline
PBE+U=0  & 5.838  & 1.400 & 9.410 & -0.103 &  0.292 & -0.259  \\
PBE+U=3  & 5.854  & 1.371 & 9.435 & -0.086 &  0.263 & -0.234  \\
PBE+U=5  & 5.864  & 1.350 & 9.451 & -0.077 &  0.242 & -0.217  \\
PBE+U=7  & 5.872  & 1.328 & 9.469 & -0.068 &  0.220 & -0.200  \\
PBE+U=11 & 5.889  & 1.279 & 9.508 & -0.052 &  0.171 & -0.161  \\ \hline
sCI-MD & 5.941  & 1.108 & 9.669 &  0.000 &  0.000 &  0.000  \\
sCI-SD & 5.935  & 1.072 & 9.713 & -0.006 & -0.036 &  0.044  \\
  \hline \hline
\end{tabular}
\caption{Ground state L\"{o}wdin populations of LiCoO$_2$ found by PBE+U compared to reference sCI-MD as well as the dominant single reference state (sCI-SD) derived from CIPSI natural orbitals.  In the rightmost columns, $\Delta(*)$ refers to the difference of a particular quantity ($*$) relative to the reference sCI-MD ground state.
}
\label{tab:lowdin_pbeu_vs_cipsi}
\end{table*}

\section{\label{sec:summary} Conclusions}

In conclusion, the methodology developed in our study, leveraging the CIPSI approach, has significantly improved the accuracy of FN-DMC simulations for ground and excited states of strongly correlated solids. This methodological advancement, exemplified in our comprehensive analysis of LiCoO$_2$, demonstrates the potential of our approach to remove the bias of a given trial wavefunction but also facilitates a deeper understanding of the material's electronic and optical properties.

Our investigation into LiCoO$_2$ reveals that, with sufficient convergence achieved through the CIPSI framework, it is possible to accurately predict band gaps that closely align with experimental observations. This represents a notable improvement over traditional DFT+U single determinant approaches, which have shown limitations in capturing the intricate electron correlation effects within strongly correlated materials. Specifically, our study illuminates the discrepancies in band gap predictions and the inadequacy of conventional methods to accurately model the excited states of LiCoO$_2$, thereby highlighting the need for a more nuanced computational strategy.

The use of CIPSI, a multi-reference selected-CI method, revealed a richer picture of the excitations in LiCoO$_2$. Incorporating multi-reference wavefunctions, our analysis reveals the involvement of multiple Kohn-Sham energy states in single-particle excitations. The inability of any single Kohn-Sham state to describe the excitation hole shows why excitations constructed via single particle-hole excitations in this Kohn-Sham basis fail to accurately describe the optical gap in LiCoO$_2$. In addition, our natural orbital analysis shows that both the ground and first excited state of the system remain well described by a single reference wavefunction, albeit one that falls outside the states produced by DFT+U.
This distinction suggests that the description of $d$-band excitations poses a challenge for PBE+U methods in this system, while natural orbitals or multi-reference approaches offer superior nodal surfaces for low-spin excitations in LiCoO$_2$. This is further supported by our findings from CIPSI analysis, where we identified the low-energy excitation as Mott-like ($t_{2g}$ to $e_g$) and dominated by a single particle-hole transition, albeit with significant static correlation contributions.

Ultimately, our findings underscore the paramount importance of the development of beyond-DFT methods to accurately describe the essential physics of excited state wavefunctions in correlated materials such as LiCoO$_2$. Our results strongly suggest that the commonly utilized single-reference DFT+U orbitals, prevalent in DMC studies of transition metal oxides, prove inadequate in capturing the complexities of excited states within $d$-band transition metal oxide systems where multiple energy states may sensitively mix to contribute to the excitation process. The superior nodal surfaces exhibited by selected-CI natural orbitals in this strongly-correlated system suggest that this method offers robust information to guide simpler single reference methods for future investigations of excited states in other correlated systems.

\section{Supporting Information}
Geometries, Inputs and Output files for DFT, DMC and sCI calculations have been made available at the Materials Data Facility\cite{blaiszik2016materials,blaiszik2019data}, {\tt https://doi.org/10.18126/z20s-pg52 }, DOI: 10.18126/z20s-pg52

\section{Acknowledgements}
Authors are grateful to Dr. Paul R. C. Kent and Pr. Lubos Mitas for very useful and enlightening conversations. H.S, K.G, J.K and A.B were supported by the U.S. Department of Energy, Office of Science, Basic Energy Sciences, Materials Sciences and Engineering Division, as part of the Computational Materials Sciences Program and Center for Predictive Simulation of Functional Materials. An award of computer time was provided by the Innovative and Novel Computational Impact on Theory and Experiment (INCITE) program. This research used resources of the Argonne Leadership Computing Facility, which is a DOE Office of Science User Facility supported under contract DE-AC02-06CH11357. We also gratefully acknowledge the computing resources provided on Bebop, a high-performance computing cluster operated by the Laboratory Computing Resource Center at Argonne National Laboratory.\\
The submitted manuscript has been created by UChicago Argonne, LLC, Operator of Argonne National Laboratory ("Argonne"). Argonne, a U.S. Department of Energy Office of Science laboratory, is operated under Contract No. DE-AC02-06CH11357. The U.S. Government retains for itself, and others acting on its behalf, a paid-up nonexclusive, irrevocable worldwide license in said article to reproduce, prepare derivative works, distribute copies to the public, and perform publicly and display publicly, by or on behalf of the Government. The Department of Energy will provide public access to these results of federally sponsored research in accordance with the DOE Public Access Plan (http://energy.gov/downloads/doe-public-access-plan).

\bibliography{main}

\end{document}